\documentclass[conference]{IEEEtran}
\newcommand{\myDots}{\ifmmode\mathinner{\ldotp\kern-0.2em\ldotp\kern-0.2em\ldotp}\else.\kern-0.13em.\kern-0.13em.\fi}
\usepackage{cite}
\usepackage{amsmath,amssymb,amsfonts}
\usepackage{algorithmic}
\usepackage{graphicx}
\usepackage{textcomp}
\usepackage{subfig}
\usepackage{fixltx2e}
\usepackage{stfloats}
\usepackage[ruled,vlined]{algorithm2e}
\usepackage{xcolor}
\usepackage{gensymb}
\usepackage{geometry}
\usepackage{adjustbox}
\def\BibTeX{{\rm B\kern-.05em{\sc i\kern-.025em b}\kern-.08em
    T\kern-.1667em\lower.7ex\hbox{E}\kern-.125emX}}
\begin{document}

\title{Phase Selection and Analysis for Multi-frequency Multi-user RIS Systems Employing Subsurfaces\\
{\footnotesize }
}
\author{
    \IEEEauthorblockN{Amy S. Inwood\IEEEauthorrefmark{2}\IEEEauthorrefmark{3}, Peter J. Smith\IEEEauthorrefmark{1}, Philippa A. Martin\IEEEauthorrefmark{2}, Graeme K. Woodward\IEEEauthorrefmark{3}} 
    
    \IEEEauthorblockA{\IEEEauthorrefmark{2}Department of Electrical and Computer Engineering, University of Canterbury, Christchurch, New Zealand }
    \IEEEauthorblockA{\IEEEauthorrefmark{3}Wireless Research Centre, University of Canterbury, Christchurch, New Zealand}
    \IEEEauthorblockA{\IEEEauthorrefmark{1}School of Mathematics and Statistics, Victoria University of Wellington, Wellington, New Zealand}

    email: amy.inwood@pg.canterbury.ac.nz, peter.smith@vuw.ac.nz, (philippa.martin, graeme.woodward)@canterbury.ac.nz,\\ 
}
\maketitle

\begin{abstract}
In this paper, we analyse the performance of a reconfigurable intelligent surface (RIS) aided system where the RIS is divided into subsurfaces. Each subsurface is designed specifically for one user, who is served on their own frequency band. The other subsurfaces (those not designed for this user) provide additional uncontrolled scattering. A new subsurface RIS design is developed based on the optimal single-user design for a pure line-of-sight (LoS) base station (BS) to RIS channel. This is also extended to arbitrary BS-RIS channels.  For our method, exact closed form solutions for the mean SNR and a mean rate upper bound are derived for the BS-RIS LoS scenario. For each user, the designed subsurface performs optimally in LoS conditions and is remarkably robust to non-LoS conditions. The system design drives down complexity to extremely low levels, reducing RIS design and receiver processing complexity and reducing the channel estimation requirements. We also quantify the complexity-performance trade-off for the new design relative to multi-user approaches.

\end{abstract}


\section{Introduction}
Reconfigurable Intelligent Surfaces (RIS) allow the manipulation of the channel between base station (BS) and users (UEs) with low power consumption. RIS consist of a grid of reflective panels (RIS elements), where each can be set to control the phase of the reflected signal. A key challenge of RIS design is selecting these phases to maximise system performance. A method for selecting the optimal phases for a single-user (SU) system where the BS to RIS channel is line-of-sight (LoS) is detailed in \cite{singh_optimal_2021}. This method is closed form and has low complexity, but it is only for the SU case. An analytically determined optimal RIS design does not appear feasible for a multi user (MU) system. Iterative schemes have been designed  \cite{singh_efficient_2022,abeywickrama_intelligent_2020, gao_robust_2021, buzzi_ris_2021}, but have high computational complexity. Among these, possibly the lowest complexity design is in \cite{singh_efficient_2022} and this is used as a benchmark in Sec. \ref{sec:numres}.

Therefore, our motivation is to design a well-performing MU RIS phase selection method with low computational complexity. In this paper, we go further and also reduce the requirements for channel estimation, a known difficulty for RIS systems, and receiver processing. For the latter, we note that simple beamforming is more practical at higher frequencies and SU systems remain of strong interest in 5G \cite{eric}. Hence, we assess the performance of a RIS system where very low system complexity is required. To do this, we treat a RIS of $N$ elements as $K$ subsurfaces with $N_1,N_2,\dots,N_K$ elements, respectively, where $K$ is the number of users. Each user will operate on a different frequency, so that matched filtering (MF) can be used at the receiver, and one subsurface is designed per user. The elements designed for other users provide uncontrolled scattering that may assist the user, particularly in higher frequency systems where there is little scattering. Therefore, we make the following contributions:
\begin{itemize}
    \item We adapt the low complexity LoS phase selection method in [1] to a MU system. We design a new sub-optimal simple phase selection method for systems with non-line-of-sight (NLoS) BS to RIS channels.
    \item We derive an exact closed form expression for the mean signal-to-noise ratio (SNR) for an uplink (UL) system with a LoS BS to RIS channel.
    \item We investigate the performance of both LoS and NLoS phase selection methods through analysis and simulations. The SNR and rate of both methods are compared to the subsurface optimal design, generated by an interior point algorithm (IPA). We also compare our methods to the existing non-subsurface MU method in \cite{singh_efficient_2022}.
    \item We compare the data rate of our methods when elements are physically grouped by user versus interleaved.
\end{itemize}
\textit{Notation}: $\mathbb{E}[\cdot]$ represents statistical expectation. $\mathbb{C}$ is the set of complex numbers. $\Re[\cdot]$ is the Real operator. $\mathcal{CN}(\boldsymbol\mu,\mathbf{Q})$ represents a complex Gaussian distribution with mean $\boldsymbol\mu$ and covariance matrix $\mathbf{Q}$. $||\cdot||$ is the Euclidean norm. ${}_2F_1(a,b,c;z)$ is the Gaussian hypergeometric function. $\otimes$ is the Kronecker product operator. $(\cdot)^T$, $(\cdot)^*$ and $(\cdot)^\dagger$ represent the transpose, conjugate and Hermitian transpose operators, respectively. The angle of a vector, $\mathbf{x}$, of length $N$ is denoted as $\angle\mathbf{x}=[\angle{x}_1,\dots,\angle{x}_N]^T$ and the exponent as $e^{\mathbf{x}}=[e^{{x}_1},\dots,e^{{x}_N}]^T$. $\lambda_\mathrm{max}(\cdot)$ and $\mathbf{v}_\mathrm{max}(\cdot)$ represent the maximum eigenvalue and eigenvector of a given matrix.
\section{System Model}\label{SysMod}
\begin{figure}[ht]
    \centering
    \includegraphics[scale=0.7]{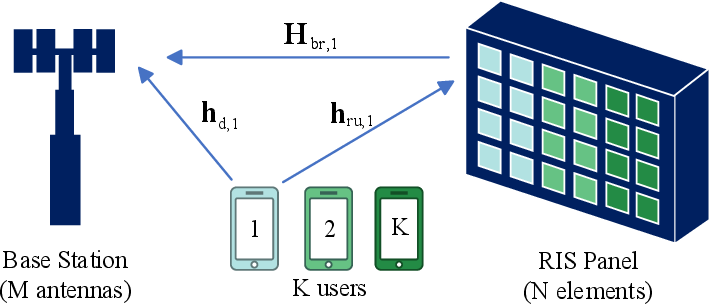}
    \caption{System model showing channels for UE 1 in band 1.}
    \label{fig:system_diagram}
\end{figure}

We consider the RIS aided system in Fig. \ref{fig:system_diagram}, where a RIS panel with $N$ elements is located near a BS with $M$ antennas. $K$ single-antenna users are located in the vicinity of the BS and RIS. The system bandwidth, $B$, is split into $K$ bands of $\frac{B}{K}$ Hz, with one user per band. The RIS phases are fixed across frequency bands. The RIS elements are grouped into ``subsurfaces'' each  designed for a different user.
\subsection{Channel Model}
For user $k$ in band $k$, let $\mathbf{h}_{\mathrm{d},k} \in \mathbb{C}^{M\times 1}$, $\mathbf{h}_{\mathrm{ru},k} \in \mathbb{C}^{N\times 1}$ and $\mathbf{H}_{\mathrm{br},k} \in \mathbb{C}^{M\times N}$ be the direct channel, the UE-RIS channel and the RIS-BS channel, respectively. $\mathbf{\Phi}\in \mathbb{C}^{N\times N}$ is a diagonal matrix of reflection coefficients which can be given in block diagonal form, where the $k$-th block is designed to enhance the channel for user $k$. $\mathbf{\Phi} = \mathrm{diag}\{\mathbf{\Phi}_1 \dots \mathbf{\Phi}_K\}$, where $\mathbf{\Phi}_k=\mathrm{diag}(e^{j\phi_{k,1}} \dots e^{j\phi_{k,N_k}})$. $N_k$ is the number of RIS elements chosen to support user $k$ and $\sum^K_{k=1}N_k=N$. The received signal in band $k$ at the BS is therefore 
\begin{equation}
    \mathbf{r}_k = (\mathbf{h}_{\mathrm{d},k}+\mathbf{H}_{\mathrm{br},k}\mathbf{\Phi h}_{\mathrm{ru}, k})s_k + \mathbf{n}_k, \label{eq:channel}
\end{equation}
where $s_k$ is the signal from user $k$, and $\mathbf{n}_k$ is additive white Gaussian noise. Using the block diagonal form, \eqref{eq:channel} becomes 
\begin{equation}    
    \begin{split}
       \mathbf{r}_k= (\mathbf{h}_{\mathrm{d},k} +&\mathbf{H}_{\mathrm{br},k,k} \mathbf{\Phi}_{k} \mathbf{h}_{\mathrm{ru},k,k})s_k \\ &+ \left(\sum_{t\neq k}^{K}(\mathbf{H}_{\mathrm{br},k,t} \mathbf{\Phi}_{t} \mathbf{h}_{\mathrm{ru},k,t}\right)s_k + \mathbf{n}_k, \label{eq:channel_split}
    \end{split} 
\end{equation}
where $\mathbf{H}_{\mathrm{br},k} = \left[\mathbf{H}_{\mathrm{br},k,1} \dots \mathbf{H}_{\mathrm{br},k,K}\right]$, $\mathbf{H}_{\mathrm{br},k,i} \in \mathbb{C}^{M\times N_i}$, $\mathbf{h}^T_{\mathrm{ru},k} = \left[\mathbf{h}^T_{\mathrm{ru},k,1} \dots \mathbf{h}^T_{\mathrm{ru},k,K}\right]$ and $\mathbf{h}_{\mathrm{ru},k,i} \in \mathbb{C}^{N_i \times 1}$.
In this channel model, we consider the correlated Rayleigh channels $\mathbf{h}_{\mathrm{d},k} = \sqrt{\beta_{\mathrm{d},k}}\mathbf{R}_{\mathrm{d},k}^{1/2}\mathbf{u}_{\mathrm{d},k}$ and $\mathbf{h}_{\mathrm{ru},k} = \sqrt{\beta_{\mathrm{ru},k}}\mathbf{R}_{\mathrm{ru},k}^{1/2}\mathbf{u}_{\mathrm{ru},k}$, where $\beta_{\mathrm{d},k}$ and $\beta_{\mathrm{ru},k}$ are the channel gains, $\mathbf{R}_{\mathrm{d},k}$ and $\mathbf{R}_{\mathrm{ru},k}$ are the correlation matrices for the UE-BS and RIS-UE links, respectively, and $\mathbf{u}_{\mathrm{d},k}, \mathbf{u}_{\mathrm{ru},k} \sim \mathcal{CN}(\mathbf{0,I})$. In this paper, we will consider both a rank-1 LoS channel and a scattered NLoS channel for $\mathbf{H}_{\mathrm{br},k}$. For the LoS case, $\mathbf{H}_{\mathrm{br},k,\mathrm{LoS}} = \sqrt{\beta_{\mathrm{br},k}}\mathbf{a}_\mathrm{b}\mathbf{a}_\mathrm{r}^\dagger$, where $\mathbf{a}_\mathrm{b}$ and $\mathbf{a}_\mathrm{r}^\dagger=[\mathbf{a}^\dagger_{\mathrm{r},1} \dots \mathbf{a}^\dagger_{\mathrm{r},K}]$ are steering vectors for the LoS ray at the BS and RIS, respectively. For the NLoS case, we adopt the Kronecker correlation model, $\mathbf{H}_{\mathrm{br},k,\mathrm{NLoS}}\!=\! \sqrt{\beta_{\mathrm{br},k}}\mathbf{R}_{\mathrm{b},k}^{1/2}\mathbf{U}_{\mathrm{br},k}\mathbf{R}_{\mathrm{r},k}^{1/2}$, where ${\mathbf{U}_{\mathrm{br},k}} \sim\mathcal{CN}(\mathbf{0,I})$, and $\mathbf{R}_{\mathrm{b},k}$ and $\mathbf{R}_{\mathrm{r},k}$ are the correlation matrices for the RIS-BS link at the BS and RIS ends, respectively.

\section{Phase Selection Methods}
\subsection{Subsurface Design for LoS Systems}
\label{sec:LOSsub}
This section considers the selection of RIS phases when $\mathbf{H}_\mathrm{br}$ is a rank-1 LoS channel. The subsurface design (SD) groups the RIS elements into $K$ subsurfaces. Each user operates on a different frequency, and each subsurface is designed independently to best serve one user. The subsurfaces not designed for user $k$ provide extra scattering and essentially become part of the SU channel not controlled by the RIS. The SD provides a number of key complexity advantages. Serving only one user in each frequency band means that simple MF is the optimal type of receiver processing. For MF, there already exists an optimal SU phase selection method as detailed in \cite{singh_optimal_2021}. Finally, channel estimation requirements are also reduced. For example, estimation of $\mathbf{h}_{\mathrm{ru},k}$, $k=1,2, \ldots ,K$, for typical RIS designs usually requires $KN$ channel elements to be estimated in any sub-band. In contrast, the SD only requires $N$ elements, ie., $N_k$ elements for user $k$. Using MF, the SNR for the section of the channel designed for user $k$, is denoted $\mathrm{SNR}_k^\mathrm{design}=\frac{E_s}{\sigma^2}\mathbf{h}_k^\dagger \mathbf{h}_k$, where $\mathbf{h}_k = \mathbf{h}_{\mathrm{d},k} +\mathbf{H}_{\mathrm{br},k,k} \mathbf{\Phi}_{k} \mathbf{h}_{\mathrm{ru},k,k}$ from (\ref{eq:channel_split}). $E_s=\mathbb{E}[|s_k|^2]$ is the transmitted signal energy and $\sigma^2=\mathbb{E}[|(\mathbf{n}_k)_i|^2]$ is the noise variance. Using the method in \cite{singh_optimal_2021}, the optimal RIS phase matrix for this channel is 
\begin{equation}
	\label{eq:Phi}
	\mathbf{\Phi}_k = \nu_k \, \mathrm{diag}\left(e^{j\angle \mathbf{a}_{\mathrm{r},k}}\right)\mathrm{diag}\left(e^{-j\angle \mathbf{h}_{\mathrm{ru},k,k}}\right),
\end{equation}
where
\begin{equation}
    \label{eq:nus}
    \nu_k ={\mathbf{a}_\mathrm{b}^\dagger\mathbf{h}_{\mathrm{d},k}}/{|\mathbf{a}_\mathrm{b}^\dagger\mathbf{h}_{\mathrm{d},k}|}.
\end{equation}
\subsection{Extended Subsurface Design for NLoS Systems}\label{sec:Phi_NLoS}
For the case when $\mathbf{H}_\mathrm{br}$ is not a rank-1 channel, there is no corresponding optimal SU design that we are aware of. Therefore, we propose adapting the method described in Sec. \ref{sec:LOSsub} into an extended subsurface design (ESD) where only the leading singular vector of $\mathbf{H}_{\mathrm{br},k,k}$ is considered. Let $\mathbf{H}_{\mathrm{br},k,k}=\mathbf{UDV}^\dagger$ be the singular value decomposition where $\mathbf{D} = \mathrm{diag}(D_{11}, D_{22}, \dots, D_{mm})$, $m = \min(M, N_k)$ and the singular values $D_{11} \geq D_{22} \geq ... \geq D_{mm}$. If $\mathbf{u}_k$ and $\mathbf{v}_k$ are the first columns of $\mathbf{U}$ and $\mathbf{V}$ respectively, then $D_{11}\mathbf{u}_k\mathbf{v}_k^\dagger$ is the strongest rank one component of $\mathbf{H}_{\mathrm{br},k,k}$. This component has the same structure as the pure LoS channel so that approximating $\mathbf{H}_{\mathrm{br},k,k}$ by $D_{11}\mathbf{u}_k\mathbf{v}_k^\dagger$ leads to the same design as in (\ref{eq:Phi}), where the steering vector $\mathbf{a}_{\mathrm{r},k}$ is replaced by $\mathbf{v}_k$. This design method is motivated and derived thoroughly in Appendix A, leading to the final result
\begin{equation}
    \label{eq:Phi_NLoS}
    \mathbf{\Phi}_\mathrm{k} = \omega_k\,\mathrm{diag}(e^{j\mathbf{\angle v}_{k}})\mathrm{diag}(e^{-j\mathbf{\angle h}_{\mathrm{ru},k}}),
\end{equation}
where
\begin{equation}
    \omega_k=\frac{|\mathbf{h}_{\mathrm{ru},k,k}|^T \mathrm{diag}(e^{-j\angle \mathbf{v}_k}) \mathbf{H}_{\mathrm{br},k,k}^\dagger \mathbf{h}_{\mathrm{d},k}}{\left||\mathbf{h}_{\mathrm{ru},k,k}|^T \mathrm{diag}(e^{-j\angle \mathbf{v}_k}) \mathbf{H}_{\mathrm{br},k,k}^\dagger \mathbf{h}_{\mathrm{d},k}\right|}.
\end{equation}
 The ESD performs close to the optimal obtained by an IPA. This will be shown in Sec. \ref{sec:perfcomp}.
 
\section{Analysis}\label{SecAnalysis}
\subsection{Expected SNR for LoS Systems Utilising the SD} \label{sec:ESNR_LoS}
The received signal $\mathbf{r}_k$ from (\ref{eq:channel_split}) can be rewritten as
\begin{equation}
\label{eq:rk}
	\mathbf{r}_k = (\mathbf{h}_{\mathrm{d},k} + \mathbf{f}_k + \mathbf{g}_k)s_k + \mathbf{n}_k,
\end{equation}
where
{
\begin{alignat}{5}
	&\mathbf{f}_k && = \mathbf{H}_{\mathrm{br},k,k}\mathbf{\Phi}_k \mathbf{h}_{\mathrm{ru},k,k} && = \sqrt{\beta_{\mathrm{br},k}}\mathbf{a}_\mathrm{b} \mathbf{a}_{\mathrm{r},k}^\dagger\mathbf{\Phi}_k \mathbf{h}_{\mathrm{ru},k,k}, \label{eq:ogf}\\
	&\mathbf{g}_k && = \sum_{t\neq k}^{K}\mathbf{H}_{\mathrm{br},k,t}\mathbf{\Phi}_t \mathbf{h}_{\mathrm{ru},k,t} && = \sum_{t\neq k}\sqrt{\beta_{\mathrm{br},k}}\mathbf{a}_\mathrm{b} \mathbf{a}_{\mathrm{r},t}^\dagger\mathbf{\Phi}_t \mathbf{h}_{\mathrm{ru},k,t}. \label{eq:ogg}
\end{alignat}}
From (\ref{eq:rk}), the SNR is given by 
\begin{equation*}
    \begin{split}
    	\mathrm{SNR}_k = \frac{E_s}{\sigma^2} & \Big[\mathbf{h}_{\mathrm{d},k}^\dagger \mathbf{h}_{\mathrm{d},k} + \mathbf{h}_{\mathrm{d},k}^\dagger \mathbf{f}_k + \mathbf{h}_{\mathrm{d},k}^\dagger \mathbf{g}_k + \mathbf{f}_k^\dagger \mathbf{h}_{\mathrm{d},k} \\ &+ \mathbf{f}_k^\dagger \mathbf{f}_k  + \mathbf{f}_k^\dagger \mathbf{g}_k + \mathbf{g}_k^\dagger \mathbf{h}_{\mathrm{d},k} + \mathbf{g}_k^\dagger \mathbf{f}_k + \mathbf{g}_k^\dagger \mathbf{g}_k\Big].
	\end{split}
\end{equation*}
Assuming that $\mathbf{h}_{\mathrm{d},k}$ and $\mathbf{h}_{\mathrm{ru},k}$ are  correlated Rayleigh, $\mathbb{E}[\mathbf{h}_{\mathrm{d},k}] = \mathbb{E}[\mathbf{h}_{\mathrm{ru},k}] = \mathbf{0}$. Using the independence of $\mathbf{h}_{\mathrm{d},k}$ and $\mathbf{h}_{\mathrm{ru},k}$ and between $\mathbf{h}_{\mathrm{ru},k}$ and $\mathbf{h}_{\mathrm{ru},t}$ for $t \neq k$, we obtain
\begin{equation}\label{meansnr}
\begin{split}
    	\mathbb{E}[\mathrm{SNR}_k] = \frac{E_s}{\sigma^2}\mathbb{E} \Big[& \mathbf{h}_{\mathrm{d},k}^\dagger \mathbf{h}_{\mathrm{d},k} + \mathbf{f}_k^\dagger \mathbf{f}_k + 2\Re(\mathbf{h}_{\mathrm{d},k}^\dagger \mathbf{f}_k)  + \mathbf{g}_k^\dagger \mathbf{g}_k \Big].
\end{split}
\end{equation}
From the work in \cite{singh_optimal_2021} on single user RIS, the direct path and user designed channel terms are 
\begin{equation}
    \mathbb{E}[\mathbf{h}_{\mathrm{d},k}^\dagger \mathbf{h}_{\mathrm{d},k}] = \beta_{\mathrm{d},k} M,
    \label{eq:hddaghd}
\end{equation}
\begin{equation}
	\mathbb{E}[\mathbf{h}_{\mathrm{d},k}^\dagger \mathbf{f}_k] = \frac{N_k\pi \sqrt{\beta_{\mathrm{d},k}\beta_{\mathrm{br},k}\beta_{\mathrm{ru},k}}\, ||\mathbf{R}_\mathrm{d}^{1/2}\mathbf{a}_\mathrm{b}||_2}{2},
	\label{eq:hddagf}
\end{equation}
\begin{equation}
	\mathbb{E}[\mathbf{f}_k^\dagger \mathbf{f}_k] = M\beta_{\mathrm{br},k}\beta_{\mathrm{ru},k}(N_k + F),
	\label{eq:fdagf}
\end{equation}
where
\begin{equation}
    F = \sum\limits_{i=1}^{N_k}\sum\limits_{j\neq i}\frac{\pi}{4}\:{}_2F_1\left(-\frac{1}{2},-\frac{1}{2};1;|\rho_{ij}|^2\right),
\end{equation}
$\rho_{ij}=(\mathbf{R}_{\mathrm{ru},k,k})_{i,j}$ and $\mathbf{R}_{\mathrm{ru},k,k}$ is the $k$th diagonal block of $\mathbf{R}_{\mathrm{ru},k}$. The impact of the subsurfaces not designed for user $k$ also needs to be considered. This is done by finding $\mathbb{E}[\mathbf{g}_k^\dagger \mathbf{g}_k]$. It is derived in Appendix B, with the result being 
\begin{equation}
\begin{split}
     	\mathbb{E}[\mathbf{g}_k^\dagger \mathbf{g}_k] = 
        M &\beta_{\mathrm{br},k}\beta_{\mathrm{ru},k}\sum\limits_{s \neq k}\sum\limits_{m=1}^{N_s}\sum\limits_{n=1}^{N_s}|r_{mn}|^2 \\ &{}_2F_1\left(\frac{1}{2},\frac{1}{2};2;|r_{mn}|^2\right),
    \label{eq:gdagg}
\end{split}
\end{equation}
where $r_{mn} = (\mathbf{R}_{\mathrm{ru},k,s})_{m,n}$ and $\mathbf{R}_{\mathrm{ru},k,s}$ is the $(k,s)$-th off-diagonal block of $\mathbf{R}_{\mathrm{ru},k}$.  From the analysis, the following insights are drawn.  In (\ref{eq:hddaghd}) - (\ref{eq:gdagg}), it is clear that increasing the number of RIS elements per user, BS elements and/or channel gain improves the SNR. Importantly, the Gaussian hypergeometric functions in (\ref{eq:fdagf}) and (\ref{eq:gdagg}) are monotonically increasing functions of the final argument. Hence, these terms increase as the spatial channel correlation is increased. The $||\mathbf{R}_\mathrm{d}^{1/2}\mathbf{a}_\mathrm{b}||$ component of (\ref{eq:hddagf}) usually increases as the correlation between BS elements is decreased. This is discussed in detail in \cite{singh_optimal_2021}. Hence, the mean SNR is assisted by obvious factors such as large system dimensions and channel gain, and by less obvious channel features, where an uncorrelated $\mathbf{h}_\mathrm{d}$ channel and a highly correlated $\mathbf{h}_\mathrm{ru}$ channel are beneficial.

\subsection{Rate Upper Bound}
\label{sec:rub}
The SNR of user $k$ determines the rate for user $k$, as $R_k = \log_2(1+\mathrm{SNR}_k)$. Using Jensen's inequality,
\begin{align}
    \label{eq:rub}
    \mathbb{E}[R_k] &= \mathbb{E}[\log_2(1+\mathrm{SNR}_k)] \leq \log_2(1+\mathbb{E}[\mathrm{SNR}_k]).
\end{align}
There is no analytical solution to determine $\mathbb{E}[\log_2(1+\mathrm{SNR}_k)]$ that we are aware of. However, when $\mathbf{H}_\mathrm{br}$ is a rank-1 LoS channel, $\mathbb{E}[\mathrm{SNR}_k]$ can be determined analytically as in Sec. \ref{sec:ESNR_LoS}, leading to the bound in (\ref{eq:rub}). When $\mathbf{H}_\mathrm{br}$ has a scattered component, $\mathbb{E}[\mathrm{SNR}_k]$ requires simulation.

\section{Numerical Results}
\label{sec:numres}
Numerical results were generated to verify the analysis above and explore the performance of the subsurface design. The channel gain values were selected based on the distance based path loss model detailed in \cite{wu_intelligent_2019}, where
\begin{equation}
    \beta =  C_0\left({d}/{D_0}\right)^{-\alpha},
\end{equation}
$D_0$ is the reference distance of 1 m, $C_0$ is the pathloss at $D_0$ (-30 dB), $d$ is the link distance in metres and $\alpha$ is the pathloss exponent ($\alpha_{\mathrm{d}} = \alpha_{\mathrm{br, NLoS}} =3.5$, $\alpha_{\mathrm{br,LoS}} = 2$ and $\alpha_{\mathrm{ru}} = 2.8$). Without loss of generality we assume $\sigma^2=1$ and $E_s$ is selected in all results except Fig.~\ref{fig:rateUE_plot} so that the mean output SNR of a SU system using SD is 5dB when the BS-RIS channel is pure LoS. For Fig.~\ref{fig:rateUE_plot} we use 0dB for the mean SNR so that the trend with increased numbers of users is clear. The correlation matrices $\mathbf{R}_\mathrm{d}$, $\mathbf{R}_\mathrm{ru}$, $\mathbf{R}_\mathrm{b}$ and $\mathbf{R}_\mathrm{r}$ can represent any  model. For simulation purposes, we use the Rayleigh fading correlation model proposed in \cite{bjornson_rayleigh_2021}, where
\begin{equation}
    \mathbf{R}_{n,m} = \mathrm{sinc} \left( 2 \,d_{mn} \right)\quad n,m = 1, \dots, L
\end{equation}
and $d_{mn}$ is the Euclidean distance between BS antennas/RIS elements $m$ and $n$, measured in wavelength units, and $L$ is the number of BS antennas/RIS elements.

For these simulations, the steering vectors $\mathbf{a}_\mathrm{b}$ and $\mathbf{a}_\mathrm{r}$ are based on the VURA model detailed in \cite{miller_analytical_2019}. Elements are equally spaced in the $x-z$ plane at both the BS and RIS. The $x$ and $z$ components of the BS steering vector are 
\begin{align*}
    \begin{split}
    \mathbf{a}_{\mathrm{b,x}} &= [1, e^{j2\pi d_b\sin(\theta_\mathrm{A})\sin(\phi_\mathrm{A})},\myDots, \\   & \qquad \quad \quad e^{j2\pi d_b(M_x-1)\sin(\theta_\mathrm{A})\sin(\phi_\mathrm{A})}]^T, 
    \end{split} \\
    \mathbf{a}_{\mathrm{b,z}} &= [1, e^{j2\pi d_b\cos(\theta_\mathrm{A})},\myDots, e^{j2\pi d_b(M_z-1)\cos(\theta_\mathrm{A})}]^T,
\end{align*}
and the $x$ and $z$ components of the RIS steering vector are
\begin{align*}
    \begin{split}
    \mathbf{a}_{\mathrm{r,x}} &= [1, e^{j2\pi d_r\sin(\theta_\mathrm{D})\sin(\phi_\mathrm{D})},\myDots, \\   & \qquad \quad \quad e^{j2\pi d_r(N_x-1)\sin(\theta_\mathrm{D})\sin(\phi_\mathrm{D})}]^T,
    \end{split}\\
    \mathbf{a}_{\mathrm{r,z}} &= [1, e^{j2\pi d_r\cos(\theta_\mathrm{D})},\myDots, e^{j2\pi d_r(N_z-1)\cos(\theta_\mathrm{D})}]^T,
\end{align*}
where $M=M_xM_z=32$, $N=N_xN_z=128$, $d_b$ and $d_r$ are the element spacings at the BS and RIS, respectively, in wavelengths, $\theta_\mathrm{D}$ and $\phi_\mathrm{D}$ are the elevation and azimuth AODs at the RIS and $\theta_\mathrm{A}$ and $\phi_\mathrm{A}$ are the corresponding elevation and azimuth AODs at the BS. Therefore, the BS and RIS steering vectors are given by 
\begin{equation}
    \mathbf{a}_\mathrm{b}= \mathbf{a}_{\mathrm{b,x}}\otimes \mathbf{a}_{\mathrm{b,z}}, \quad\quad\quad \mathbf{a}_\mathrm{r}= \mathbf{a}_{\mathrm{r,x}}\otimes \mathbf{a}_{\mathrm{r,z}},
\end{equation}
respectively. We assume the RIS is on a $\frac{\pi}{4}$ angle with respect to the BS, so $\phi_\mathrm{D} = \frac{5\pi}{4}$ and $\phi_\mathrm{A} = \frac{\pi}{4}$. We also assume they are at the same height, so $\theta_\mathrm{D}=\theta_\mathrm{A}=\frac{\pi}{2}$. For all simulations, $10^4$ replicates were generated. These parameter values and definitions do not change throughout the results, unless specified otherwise. 

\subsection{Performance Comparison of Phase Selection Methods}
\label{sec:perfcomp}
\begin{figure}[h!]
    \centering
    \vspace{-0.5em}
    \includegraphics[scale=0.47]{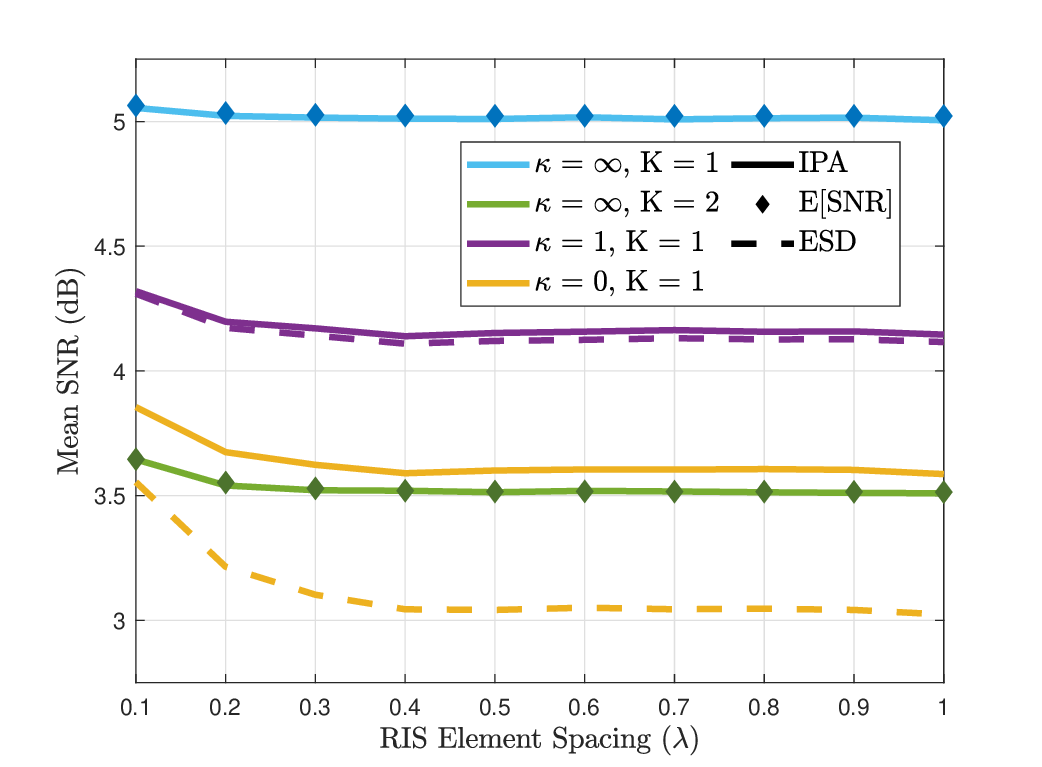}
    \caption{Mean SNR comparison between the $\mathbb{E}[\mathrm{SNR}_1]$ from (\ref{meansnr}), the ESD method from Sec. \ref{sec:Phi_NLoS} and the IPA optimal.}
    \label{fig:ESNR_plot}
\end{figure}

Fig. \ref{fig:ESNR_plot} verifies the analytical results and investigates the SD and ESD methods for four parameter sets. The amount of scattering in $\mathbf{H}_\mathrm{br}$ is adjusted with the K-factor, $\kappa$. The $\kappa=\infty$ curves compare the analytical $\mathbb{E}[\mathrm{SNR}_1]$ from Sec. \ref{sec:ESNR_LoS} with the SNR generated by maximising (\ref{meansnr}) for both $K=1$ and $K=2$. An IPA was used to search for the optimal RIS phases for each user. These curves verify the analysis for LoS links in Sec. \ref{sec:ESNR_LoS} and the optimal SU phase design in \cite{singh_optimal_2021}. When $\kappa= 1$, $\mathbf{H}_\mathrm{br}$ is partially LoS and partially scattered. Phases are again selected using the ESD and the IPA. The ESD performance is remarkably close to the IPA, showing that it is near optimal even when the scattered component has the same strength as the LoS link. The $\kappa = 0$ curves show the simulated mean SNR when $\mathbf{H}_\mathrm{br}$ is fully scattered, the ultimate test of robustness to the LoS assumption. Phases are selected using the ESD and the IPA. The ESD is clearly sub-optimal, but performs between $84-90\%$ of the IPA over the full range of $d_r$. 

It is clear from Fig. \ref{fig:ESNR_plot} that decreasing the distance between RIS elements increases the mean SNR. This is largely due to the increase in correlation between RIS elements, as predicted by the analysis in Sec. \ref{sec:ESNR_LoS}. When $\mathbf{H}_\mathrm{br}$ is LoS, there is no scattered component, so correlation has less effect. In contrast, when $\mathbf{H}_\mathrm{br}$ is purely scattered,  correlation effects are more important and boost the SNR.

\subsection{CDF of SNR for a Range of $K$ Values}
\begin{figure}[h!]
    \centering
    \includegraphics[scale=0.5]{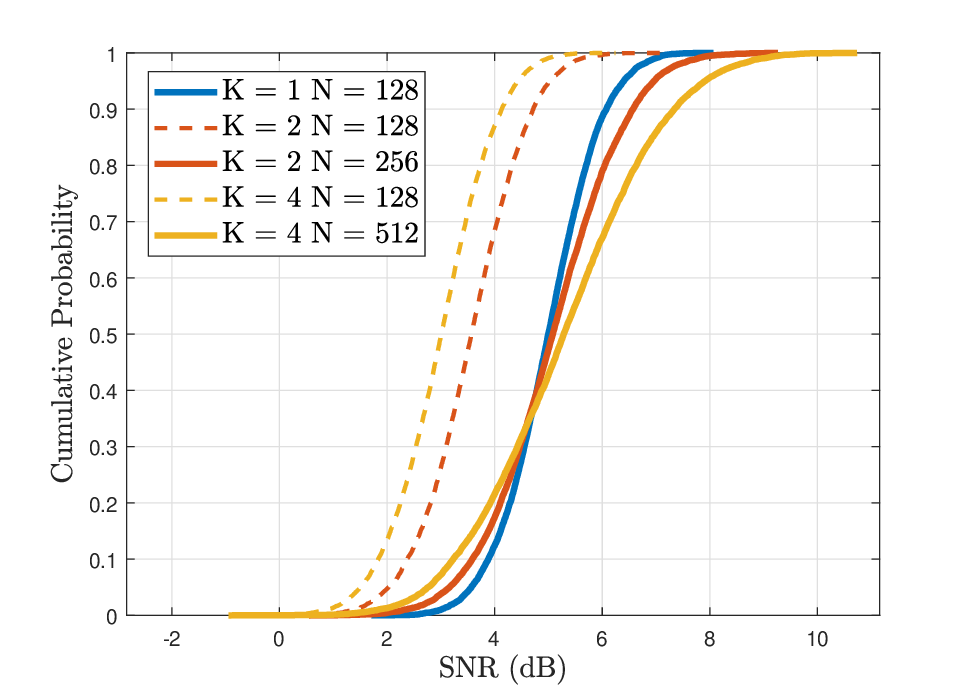}
    \caption{Simulated CDFs for $N = K\times 128$ for $K=\{1,2,4\}$.}
    \label{fig:SNRCDFs_plot}
    \vspace{-1em}
\end{figure}

Fig. \ref{fig:SNRCDFs_plot} shows the impact of $N$ and $K$ on the SNR distribution for one user. $\mathbf{H}_\mathrm{br}$ is LoS, and the LoS phase selection method was used. The solid lines show $N=128K$ total RIS elements, while the dashed lines show $N = 128$ total RIS elements. The median SNR decreases per user as more users are added if $N$ is not adjusted accordingly, as fewer elements are used to assist each user. However, if $N$ scales with $K$, an increase in user numbers is beneficial as there are more uncontrolled scatterers while the size of the designed subsurface is fixed. As more users are added, outage probabilities are increased as the SNR distribution becomes more variable due to the extra uncontrolled scattering.

\subsection{Rate Approximation Results}
\begin{figure}[t!]
    \centering
    \includegraphics[scale=0.45]{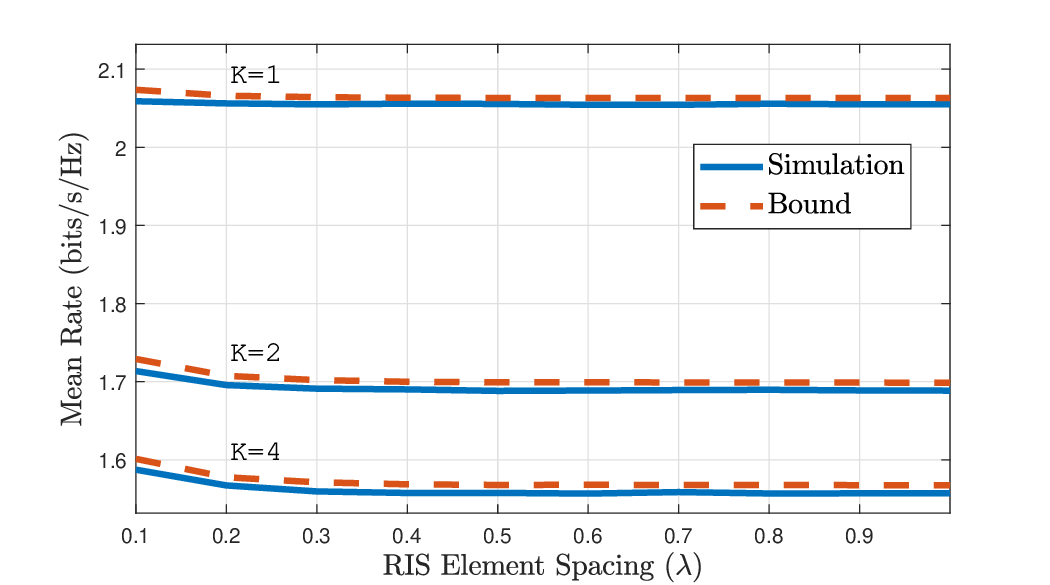}
    \caption{Mean rate vs upper bound comparison for user 1 with $K$ = \{1,2,4\} total users, $N=128$.}
    \label{fig:ratebound_plot}
    \vspace{-1em}
\end{figure}

As discussed in (\ref{eq:rub}) of Sec. \ref{sec:rub}, Jensen's inequality can be used to provide an upper bound for the expected rate. Fig. \ref{fig:ratebound_plot} compares expected rate bounds with the mean simulated rate for one user when $\mathbf{H}_{\mathrm{br}}$ is LoS.

It can be seen that the upper bound on the expected rate is very tight, particularly for increased element spacing. Less spacing between elements leads to more deviation from the bound which is due to increased variability, as explained below. At very low spacings, the channels at the RIS are highly correlated, so a good channel can be highly beneficial while a bad channel is highly damaging. In contrast, for larger spacings, correlation drops and the effects of good and bad channels tend to average out over the $N$ elements, giving more stable, less variable results. For example, the SNR variance at $0.1\lambda$ spacing is 15 times greater than at $0.4\lambda$ for $K=1$. Since the first order correction to Jensen's inequality is a negative term scaled by the variance, we see that the overestimation for small spacings is explained by increased variance. For spacings greater than $0.4\lambda$, the elements are sufficiently separated for their correlation to be so low that a similar rate is achieved.

\subsection{Impact of the Number of Users on Rate}
\begin{figure}[h!]
    \centering
    \includegraphics[scale=0.5]{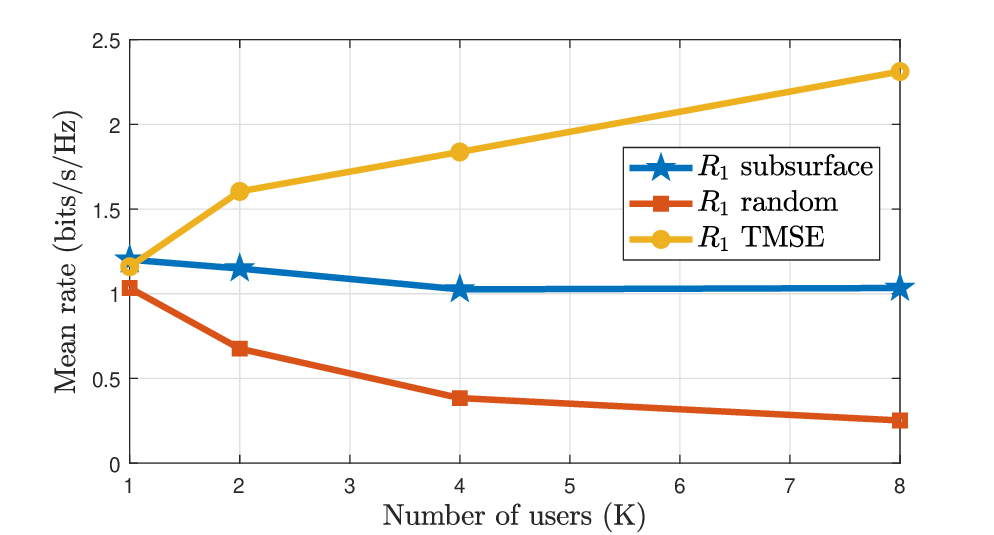}
    \caption{A comparison of ${R}_\mathrm{sum}$ for the subsurface, TMSE and random RIS phase selection methods for a range of $K$.}
    \label{fig:rateUE_plot}
    \vspace{-0.5em}
\end{figure}

 Fig. \ref{fig:rateUE_plot} gives a comparison between the SD, random phases and  the  MU approach in \cite{singh_efficient_2022} which reduces complexity by minimizing the total mean squared error (TMSE) of an MMSE receiver. The random and TMSE methods place all users in one band, so each user has $K$ times the bandwidth. This fundamental bandwidth advantage causes the growth in rate observed for the TMSE approach. In contrast, the SD decays slightly as each user has fewer RIS elements designed for it. Since SD is optimal for $K=1$, it has the highest initial value. As expected, the random designs give a lower bound, which is close to SD and TMSE at $K=1$ as the SNR is set to a low value making the direct channel dominant here.
 
 Clearly, the SD cannot hope to deliver similar rates to MU approaches due to the bandwidth restriction. Nevertheless, as shown in Sec. \ref{sec:ESNR_LoS}, SD improves with higher correlation. This suggests that  Fig.~\ref{fig:rateUE_plot} is close to the  worst case for SD, as the correlation model in \cite{bjornson_rayleigh_2021} results in low correlation values. Results, not shown here for reasons of space, show that Ricean channels and increased correlation enhance the performance of SD and close the gap to the TMSE design. 
 
 The trade-off for the rate loss is substantial complexity reduction, improved fairness and reduced channel estimation:
\begin{itemize}
\item \textbf{RIS design:} Here, we evaluate the complexity of the TMSE design and SD, based on the number of complex multiplies and the complexity of any matrix operations. Basic implementations of both techniques require
\begin{align}\textrm{TMSE:}& \, N(K\!+\!K^2)\!+\!M(K\!+\!K^2)\!+\!K(3\!+\!3K+\!2K^2)\notag\\
\textrm{SD:}& \, 2N+MK \notag
\end{align}
complex multiplies. Hence, SD provides a substantial complexity reduction of a factor of $(K+K^2)/2$ assuming $N$ is the largest dimension. Furthermore, the SD requires no matrix operations, while the TMSE method requires two $K \times K$ matrix inversions and a  $K \times K$ matrix  eigen-decomposition. The subsurface method also provides fairer results according to Jain's fairness test, with an average of 0.563 compared to an average of 0.423 for TMSE for $K=8$. This is due to all users having an equal section of the RIS designed specifically for them, in comparison to overall maximisation methods that may maximise a metric by focusing on one UE.
\item \textbf{Receiver design:} The TMSE method requires MMSE processing at the receiver, which requires an additional  $K\times K$ matrix inverse operation of order $\mathcal{O}(K^3)$ relative to the MF used in the subsurface method.
\item \textbf{CSI estimation:} A UE must acquire all $N$ RIS channels for the TMSE method. For SD, the same UE only needs the $\frac{N}{K}$  RIS channels specifically designed to serve it. This gives a ${K}$-fold reduction in channel estimation.
\end{itemize}
\vspace{-0.5em}
\subsection{RIS Element Layout Results}
\begin{figure}[h!]
    \centering 
    \includegraphics[scale=0.6]{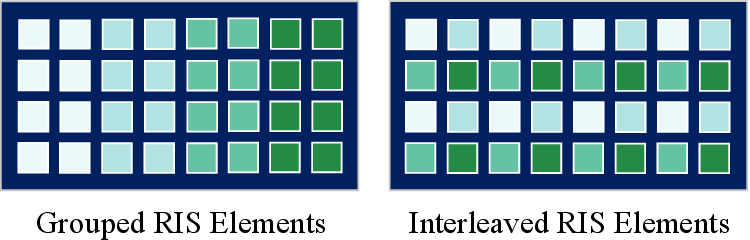}
    \caption{RIS elements allocated to each user when elements are grouped (left) and interleaved (right).}
    \label{fig:interleaving}
        \vspace{-1em}
\end{figure}

\begin{figure}[h!]
    \centering
    \includegraphics[scale=0.55]{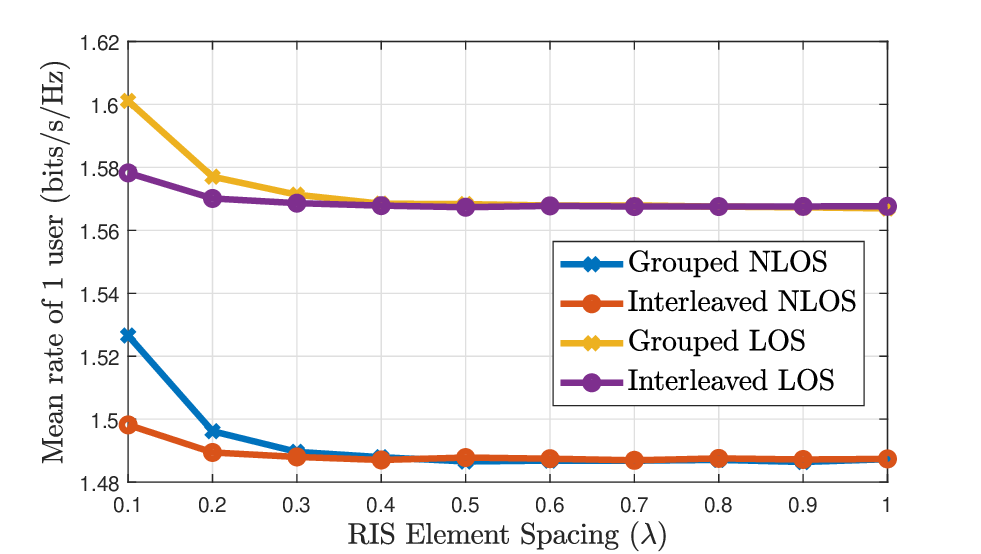}
    \caption{The user 1 rate of LoS and NLoS $\mathbf{H}_\mathrm{br}$ channels for grouped and interleaved RIS layouts and element spacing.}
    \label{fig:interleaving_plot}
\end{figure}

A key part of the subsurface design is the layout.  Fig. \ref{fig:interleaving} shows two ways of arranging the RIS elements - grouping them in blocks (the default in the system description in Secs.~\ref{SysMod}-\ref{SecAnalysis}), and placing the largest equal gap between elements serving one user (termed ``interleaved"). The single-user rate in four situations was determined through simulation using an IPA for RIS design. Grouped and interleaved elements with $\mathbf{H}_\mathrm{br}$ set as both LoS ($\kappa = \infty$) and fully scattered ($\kappa = 0$) were investigated. The rate was determined for a range of element spacings and is shown in Fig. \ref{fig:interleaving_plot}.

The single-user rate with elements arranged in blocks was 1.5\% and 1.9\% higher than the interleaved case when element spacing was 0.1$\lambda$ and $\mathbf{H}_\mathrm{br}$ is LoS and NLoS, respectively. This is explained by the analysis in Sec. \ref{sec:perfcomp} which shows that correlation helps SNR. Hence, we reach the slightly counter-intuitive conclusion that locating elements closer together is better and grouping outperforms interleaving.

\section{Conclusion}
We have proposed a MU RIS system with extremely low complexity compared to established methods. To achieve this, we design subsurfaces of the RIS for each user, with each user served on a separate frequency. This lowers the complexity of RIS design and receiver processing as well as the need for channel estimation. The mean SNR is derived, leading to insights into system performance. Numerical results quantify the performance trade off relative to MU approaches. We detail possible physical element layouts, leading to an understanding that decreasing the distance between RIS elements and thus increasing correlation is the most efficient deployment option.

\vspace{1em}
\section*{Appendix A \\$\mathbf{\Phi_k}$ derivation for NLoS channels in Sec. \ref{sec:Phi_NLoS}}
\label{sec:AppenA}
The SNR for user $k$ is given by 
\begin{equation}
    \mathrm{SNR}_k = ||\mathbf{\bar h}_k||^2\frac{E_s}{\sigma^2},
\end{equation}
where $\mathbf{\bar h}_k = \mathbf{h}_{\mathrm{d},k} + \mathbf{H}_{\mathrm{br},k,k}\mathbf{\Phi}_k \mathbf{h}_{\mathrm{ru},k,k}+\mathbf{g}_k$ is the channel for a single user from (\ref{eq:channel_split}). Squaring this gives 
\begin{equation}
\begin{split}
    ||\mathbf{\bar h}_k||^2 =& \mathbf{h}_{\mathrm{d},k}^\dagger \mathbf{h}_{\mathrm{d},k} + 2\Re\{\mathbf{h}_{\mathrm{d},k}^\dagger \mathbf{H}_{\mathrm{br},k,k}\mathbf{\Phi}_k \mathbf{h}_{\mathrm{ru},k,k}\} \\& +||\mathbf{H}_{\mathrm{br},k,k}\mathbf{\Phi}_k \mathbf{h}_{\mathrm{ru},k,k}||^2+G,
\end{split}
\end{equation}
where $G$ represents all terms involving $\mathbf{g}_k$. Neglecting terms in $\mathbf{g}_k$ which are controlled for the other users, $||\mathbf{\bar h}_k||$ is dominated by the quadratic terms, $\mathbf{h}_{\mathrm{d},k}^\dagger\mathbf{h}_{\mathrm{d},k}$ and $||\mathbf{H}_{\mathrm{br},k,k}\mathbf{\Phi}_k \mathbf{h}_{\mathrm{ru},k,k}||^2$. Therefore, we propose to use $\mathbf{\Phi}_k = \omega_k\mathbf{\Psi}_k$, where $\omega_k$ is a rotation to align $\mathbf{H}_{\mathrm{br},k,k}\mathbf{\Phi}_k \mathbf{h}_{\mathrm{ru},k,k}$ with $\mathbf{h}_{\mathrm{d},k}$. To achieve this,
\begin{equation}
    \label{eq:general_v}
    \omega_k = \frac{\mathbf{h}_{\mathrm{ru},k,k}^\dagger\mathbf{\Psi}_k^* \mathbf{H}_{\mathrm{br},k,k}^\dagger \mathbf{h}_{\mathrm{d},k}}{|\mathbf{h}_{\mathrm{ru},k,k}^\dagger\mathbf{\Psi}_k^* \mathbf{H}_{\mathrm{br},k,k}^\dagger \mathbf{h}_{\mathrm{d},k}|}.
\end{equation}
This rotation maximises the cross product $\Re\{\mathbf{h}_{\mathrm{d},k}^\dagger \mathbf{H}_{\mathrm{br},k,k}\mathbf{\Phi}_k \mathbf{h}_{\mathrm{ru},k,k}\}$ for a given $\mathbf{\Psi}_k$. To select a $\mathbf{\Psi}_k$ that maximises $||\mathbf{H}_{\mathrm{br},k,k}\mathbf{\Phi}_k \mathbf{h}_{\mathrm{ru},k,k}||^2$, let 
\begin{equation}
    \label{eq:Psi}
    \mathbf{\Psi}_k = \mathbf{A}\, \mathrm{diag}\left(e^{j\angle(\mathbf{h}_{\mathrm{ru},k,k})_1^*} \dots e^{j\angle(\mathbf{h}_{\mathrm{ru},k,k})_{N_k}^*}\right),
\end{equation}
where $\mathbf{A} = \mathrm{diag}\left(a_1 \dots a_N\right)$. This leads to
\begin{align*}
    ||\mathbf{H}_{\mathrm{br},k,k}&\mathbf{\Psi}_k \mathbf{h}_{\mathrm{ru},k,k}||^2 \\ &= |\mathbf{h}_{\mathrm{ru},k,k}|^T\mathbf{A^* H}_{\mathrm{br},k,k}^\dagger \mathbf{H}_{\mathrm{br},k,k}\mathbf{A|h}_{\mathrm{ru},k,k}| \\
    &= ||\mathbf{h}_{\mathrm{ru},k,k}||^2\,\mathbf{\Tilde{c}^\dagger H}_{\mathrm{br},k,k}^\dagger \mathbf{H}_{\mathrm{br},k,k} \mathbf{\Tilde{c}},
\end{align*}
where $\mathbf{c=A\,|h}_{\mathrm{ru},k,k}|$ and $\mathbf{\Tilde{c}=\frac{c}{||c||}}$. From the Rayleigh-Ritz Theorem
\begin{equation}
    \mathbf{\Tilde{c}^\dagger H}_{\mathrm{br},k}^\dagger \mathbf{H}_{\mathrm{br},k} \mathbf{\Tilde{c}}\, \leq \,\lambda_{\mathrm{max}}(\mathbf{H}_{\mathrm{br},k}^\dagger \mathbf{H}_{\mathrm{br},k}).
\end{equation}
This maximum is achieved when $\mathbf{\Tilde{c}} = \mathbf{v}_{\mathrm{max}}(\mathbf{H}_{\mathrm{br},k,k}^\dagger \mathbf{H}_{\mathrm{br},k,k})=\mathbf{v}_k$, where $\mathbf{v}_k$ is also the leading singular vector of $\mathbf{H}_{\mathrm{br},k,k}$. However, the RIS cannot change the amplitude of the signal. In order for $\mathbf{\Tilde{c}}$ to be as close to the maximum eigenvector as possible, $\mathbf{\Tilde{c}}$ is selected such that $|\mathbf{\Tilde{c}^\dagger v}_k|$ is maximised. This can be expanded to
\begin{equation*}
    |\mathbf{\Tilde{c}^\dagger v}_k| = \left|\mathbf{\frac{c^\dagger}{||c||}}\mathbf{v}_k\right|= \frac{|\mathbf{h}_{\mathrm{ru},k,k}|^\dagger \mathbf{A}^* \mathbf{v}_k}{||\mathbf{h}_{\mathrm{ru},k,k}||} \, \leq \, \frac{|\mathbf{h}_{\mathrm{ru},k,k}|^\dagger |\mathbf{v}_k|}{||\mathbf{h}_{\mathrm{ru},k,k}||}.
\end{equation*}
Therefore, the optimal choice is $\mathbf{A} = e^{j \angle \mathbf{v}_k}$. Combining this choice of $\mathbf{A}$ with (\ref{eq:Psi}), and substituting $\mathbf{\Psi}_k$ into (\ref{eq:general_v}), the value of $\mathbf{\Phi}_k$ is as in (\ref{eq:Phi_NLoS}).

\section*{Appendix B \\ Derivation of $\mathbb{E}[\mathbf{g}_k^\dagger \mathbf{g}_k]$}
\label{sec:AppenB}
From the definition of $\mathbf{g}_k$ in (\ref{eq:ogg}),
\begin{equation}
    	\mathbf{g}_k^\dagger \mathbf{g}_k = \beta_{\mathrm{br},k} M \sum_{s \neq k} \sum_{t \neq k} \mathbf{h}_{\mathrm{ru},k,s}^\dagger\mathbf{\Phi}_s^\dagger \mathbf{a}_{\mathrm{r},s}\mathbf{a}_{\mathrm{r},t}^\dagger\mathbf{\Phi}_t\mathbf{h}_{\mathrm{ru},k,t},
\end{equation}
using $\mathbf{a}_\mathrm{b}^\dagger \mathbf{a}_\mathrm{b} = M$. Note that the $\mathbf{h}_{\mathrm{ru},k,s}$ terms are for user $k$, while the $\mathbf{\Phi}_s$ terms are for all other users, so they are independent. Also, $\mathbb{E}[\mathbf{\Phi}_s] = 0$ as $\mathbb{E}[\nu_s] = 0$, since $\nu_s$ is the phase of a circular zero mean complex Gaussian variable from (\ref{eq:nus}). Hence when taking the expectation of $\mathbf{g}_k^\dagger \mathbf{g}_k$, only the terms where $s = t$ are non-zero. Therefore,
\begin{align}
	\mathbb{E}[\mathbf{g}_k^\dagger \mathbf{g}_k]  &=  M\beta_{\mathrm{br},k}\mathbb{E}\left[\sum_{s \neq k} \mathbf{h}_{\mathrm{ru},k,s}^\dagger\mathbf{\Phi}_s^\dagger \mathbf{a}_{\mathrm{r},s} \mathbf{a}_{\mathrm{r},s}^\dagger\mathbf{\Phi}_s \mathbf{h}_{\mathrm{ru},k,s}\right]. \notag\\
	\intertext{$\mathbb{E}[\mathbf{h}_{\mathrm{ru},k}\mathbf{h}_{\mathrm{ru},k}^\dagger] = \beta_{\mathrm{ru},k}\mathbf{R}_{\mathrm{ru},k}$, so taking the $s^{\mathrm{th}}$ block of $\mathbf{h}_{\mathrm{ru},k}$, we have $\mathbb{E}[\mathbf{h}_{\mathrm{ru},k,s}\mathbf{h}_{\mathrm{ru},k,s}^\dagger] = \beta_{\mathrm{ru},k}\mathbf{R}_{\mathrm{ru},k,s}$. Therefore,}
	\mathbb{E}[\mathbf{g}_k^\dagger \mathbf{g}_k] &= M\beta_{\mathrm{br},k}\beta_{\mathrm{ru},k}\mathrm{tr}\left\{\sum_{s\neq k}\mathbb{E}\left[\mathbf{\Phi}_s^\dagger \mathbf{a}_{\mathrm{r},s} \mathbf{a}_{\mathrm{r},s}^\dagger\mathbf{\Phi}_s\right]\mathbf{R}_{\mathrm{ru},k,s}\right\} \label{eq:eog_simple}
\end{align}
The $(m,n)$-th element of $\mathbb{E}\left[\mathbf{\Phi}_s^\dagger \mathbf{a}_{\mathrm{r},s} \mathbf{a}_{\mathrm{r},s}^\dagger\mathbf{\Phi}_s\right]$ has mean value 
\begin{equation}
    \begin{split}
        \label{eq:eoa}
        \mathbb{E}[e^{-j\phi_{s,m}}(\mathbf{a}_{\mathrm{r},s})_{m}(\mathbf{a}&_{\mathrm{r},s})_{n}^\ast e^{j\phi_{s,n}}] = \\ &(\mathbf{a}_{\mathrm{r},s})_{m}(\mathbf{a}_{\mathrm{r},s})_{n}^\ast\mathbb{E}[e^{-j\phi_{s,m}}e^{j\phi_{s,n}}].
    \end{split}
\end{equation}
From the definition of the RIS phases in (\ref{eq:Phi}),
\begin{equation}
    \label{eq:Eee}
    \begin{split}
        \mathbb{E}[&e^{j\phi_{s,n}}e^{-j\phi_{s,m}}] = (\mathbf{a}_{\mathrm{r},s})_{n}(\mathbf{a}_{\mathrm{r},s})_{m}^\ast M_{m,n} \\ &=(\mathbf{a}_{\mathrm{r},s})_{n}(\mathbf{a}_{\mathrm{r},s})_{m}^\ast\mathbb{E}\left[e^{-j\angle (\mathbf{h}_{\mathrm{ru},s,s})_{n}}e^{j\angle (\mathbf{h}_{\mathrm{ru},s,s})_{m}}\right].
    \end{split}
\end{equation}
From (4.26) of \cite{miller_complex_1974}, and letting $r_{mn} = (\mathbf{R}_{\mathrm{ru},k,s})_{m,n}$,
\begin{equation}
    M_{m,n}=\frac{\pi}{4}r_{mn}(1-|r_{mn}|^2)\,{}_2F_1 \left(1.5, 1.5, 2; |r_{mn}|^2\right).
\end{equation}
Using the hypergeometric function linear transformation specified in 15.8.1 of \cite{olver_NIST_2010},
\begin{equation}
    \label{eq:miller_us}
    M_{m,n} = \frac{\pi}{4}r_{mn}\,{}_2F_1 \left(0.5, 0.5, 2; |r_{mn}|^2\right).
\end{equation}
Combining (\ref{eq:eog_simple})-(\ref{eq:Eee}) and  (\ref{eq:miller_us}) gives (\ref{eq:gdagg}). \vspace{-0.3em} 

\bibliographystyle{IEEEtran}
\bibliography{IEEEabrv, conference.bib}
\vspace{12pt}

\end{document}